\documentclass[floatfix,aps,pre,reprint]{revtex4-2}
\usepackage{palatino}
\usepackage{graphicx}
\usepackage{amssymb}
\usepackage{amsmath}
\usepackage[hidelinks]{hyperref}
\usepackage{cleveref}
\usepackage{bm}
\usepackage{lipsum} 
\usepackage{float}
\usepackage{mathrsfs}
\usepackage{mathtools}
\usepackage{tikz}
\usepackage{pgfplots}
\usepackage{framed}
\usepackage{mdframed}
\usepackage{xcolor}
\usepackage{todonotes}
\usepackage{url}
\usepackage{physics}
\usepackage{pgfplots}

\usetikzlibrary{calc}
\usetikzlibrary{matrix}

\newcommand\ddfrac[2]{\frac{\displaystyle #1}{\displaystyle #2}}
\DeclareMathOperator*{\E}{\mathbb{E}}
\makeatletter
\Crefname{figure}{Figure}{Figures}
\crefname{figure}{Fig.}{Figs.}
\Crefname{equation}{Equation}{Equations}
\crefname{equation}{Eq.}{Eqs.}
\Crefname{section}{Section}{Sections}
\crefname{section}{Sec.}{Secs.}
\newcommand\changes[1]{{#1}}

\begin{document}
\title{The planted directed polymer: inferring a random walk from noisy images} 
\author{Sun Woo P. Kim}
\email{swk34@cantab.ac.uk}
\affiliation{Department of Physics, King's College London, Strand, London WC2R 2LS, United Kingdom}
\author{Austen Lamacraft}
\affiliation{TCM Group, Cavendish Laboratory, University of Cambridge, Cambridge CB3 0HE, United Kingdom}

\begin{abstract}
    We introduce and study the planted directed polymer, in which the path of a random walker is inferred from noisy `images' accumulated at each timestep.
    Formulated as a nonlinear problem of Bayesian inference for a hidden Markov model, this problem is a generalisation of the directed polymer problem of statistical physics, coinciding with it in the limit of zero signal to noise.
    For a 1D walker we present numerical investigations and analytical arguments that no phase transition is present. When formulated on a Cayley tree, methods developed for the directed polymer are used to show that there is a transition with decreasing signal to noise where effective inference becomes impossible, meaning that the average fractional overlap between the inferred and true paths falls from one to zero.
    
    
\end{abstract}

\maketitle



\section{Introduction}\label{section:introduction}

Recent years have seen a great deal of research activity at the interface of statistical physics and Bayesian inference \cite{zdeborova2016statistical}. Broadly speaking, the connection is as follows. Suppose the system of interest is described by some random variables $\mathbf{x}$ following a \emph{prior} probability distribution $p(\mathbf{x})$. We learn about $\mathbf{x}$ by measuring some other random variables $\mathbf{y}$, which are described by some conditional distribution or `measurement model' $p(\mathbf{y}|\mathbf{x})$ \footnote{Note that throughout we adopt the convention that different distributions are distinguished by the names of their arguments: a convenient abuse of notation}. From the values of $\mathbf{y}$ we compute the \emph{posterior distribution} $p(\mathbf{x}|\mathbf{y})$ of $\mathbf{x}$ using Bayes' rule
\begin{equation}\label{eq:bayes}
    p(\mathbf{x}|\mathbf{y}) = \frac{p(\mathbf{y}|\mathbf{x})p(\mathbf{x})}{p(\mathbf{y})},
\end{equation}
where the distribution of the measurement outcomes $p(\mathbf{y}) = \sum_{\mathbf{x}} p(\mathbf{y} \vert \mathbf{x}) p(\mathbf{x})$ can be regarded as a normalizing factor or partition function in the language of statistical physics. \changes{This setting assumes a perfect knowledge of the prior distribution and measurement model, an idealisation which could be approached in certain (highly engineered) settings, ex. transmission over a noisy channel}. Nevertheless, the computation of $p(\mathbf{x}|\mathbf{y})$ will in general be intractable when the number of variables $\mathbf{x}$ is large, due to the difficulty of calculating the denominator in \cref{eq:bayes}. The intractability of the partition function is a general feature of statistical mechanical systems, and this represents the first point of contact between statistical physics and inference. Furthermore, there are `natural' examples of inference problems \changes{in the `thermodynamic limit' of } many variables where\changes{, for certain cases,} the phenomenology of phase transitions in the thermodynamic limit of random systems is applicable, with the values of the measurements $\mathbf{y}$ corresponding to the randomness. In the inference setting, such transitions represent a change in the structure of the posterior as a parameter representing the strength of the `signal' of $\mathbf{x}$ in the measurements $\mathbf{y}$ is varied. \changes{In cases where there exists a phase transition,} as the signal weakens, the probability of successful inference goes to zero at the phase transition.

\begin{figure}[h] \centering
    \begin{tikzpicture}
        \matrix[matrix of math nodes,column sep=2em,row
        sep=2em,cells={nodes={rectangle,draw,minimum width=4.5em,minimum height=2em, inner sep=0pt, rounded corners=0.5em}},
        column 1/.style={nodes={rectangle,draw=none,minimum width=2em,minimum height=2em}},
        column 5/.style={nodes={rectangle,draw=none,minimum width=2em,minimum height=2em}}] (m) {
        \cdots & \mathbf{x}(t-1) & \mathbf{x}(t) & \mathbf{x}(t+1) & \cdots \\
         & \mathbf{y}(t-1) & \mathbf{y}(t) & \mathbf{y}(t+1) & \\
        };
        \foreach \X in {1,2,3,4}
        {\draw[-latex] (m-1-\X) -- (m-1-\the\numexpr\X+1) node[midway,above]{};};
        \foreach \X in {2,3,4}
        {\draw[-latex] (m-1-\X) -- (m-2-\X) node[pos=0.6,left]{};};
        \end{tikzpicture} \caption{Inference in a hidden Markov model}
        \label{fig:markov}
\end{figure}
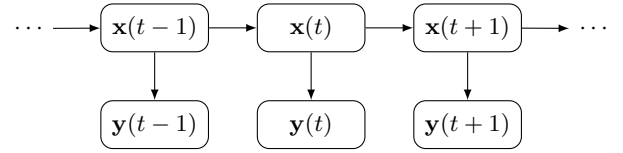

In this work, we will be concerned with developing the parallels between statistical mechanics and inference within a particular class of problems that have the structure of a \emph{hidden Markov model} (HMM) (\cref{fig:markov}) \cite{rabiner1986introduction,rabiner1989tutorial}. \changes{In an HMM, the state of the system $x_t$ follows a Markov process,}
\begin{align} \label{eq:markov}
    p(x_{1:t}) = \left(\prod_{\tau=2}^t p(x_\tau \vert x_{\tau-1})\right) p(x_1),
\end{align}
where we have introduced the notation $x_{1:t}$ to denote a sequence of variables ${x_1, \ldots x_t}$, and $p(x_\tau \vert x_{\tau-1})$ are the transition probabilities (or kernel) of the process. Measurements $y_t$ are conditional on the value $x_t$ at the same timestep through some measurement model $p(y_t|x_t)$. 

Inference in HMMs may involve one of several related tasks. \emph{Filtering} refers to the problem of obtaining the posterior $p(x_t|y_{1:t})$ of the present value conditioned on the history of measurements, while \emph{smoothing} involves conditioning additionally over future measurements. The most famous example of filtering in a HMM is the \emph{Kalman filter} \cite{welch1995introduction}, in which both $p(x_\tau \vert x_{\tau-1})$ and $p(x_t|y_t)$ are Gaussian with linear dependence on the conditioning variable. Because the product of Gaussians is also Gaussian, the filtering distribution $p(x_t|y_{1:t})$ --- obtained from Bayes' rule \cref{eq:bayes} --- is a Gaussian with an explicit form.  

Here we are concerned with a fundamentally \emph{nonlinear} observation model, consisting of `images' containing `pixels', one for each possible \changes{location} of the system $x$. \changes{This is a very natural model of tracking an object from a time series of images taken against a noisy background, ex. tracking molecules from fluorescence imaging \cite{smal2006bayesian}. From now on, we denote observations as $\phi_{x,t}$, indexed by both $x$ and $t$. They take the form}
\begin{align}\label{eq:measurement_model_intro}
    \phi_{x, t} = \psi_{x, t} + \epsilon \delta_{x, x_t},
\end{align}
where $\psi_{x,t}$ is Gaussian white noise, $\epsilon$ is a parameter that controls the relative strength of the signal (i.e. the signal-to-noise ratio), \changes{and $\delta_{x, x_t}$ is the Kronecker delta between pixel location $x$ and current walker location $x_t$.} In \cref{eq:measurement_model_intro} the shift of the mean of $\phi_{x_t,t}$ by $\epsilon$ can reveal through which pixel the system has passed. 

Problems of this type have appeared before in the engineering literature applied to tracking of paths through a noisy or cluttered environment \cite{yuille2000fundamental,offer2018phase}. Our aim is to introduce and analyse a particularly simple version of the problem that we call the \emph{planted directed polymer problem}, because of its connection with the statistical mechanics of a directed polymer in a random medium \cite{kardar2007statistical}.

A second and quite different motivation for our work is the recently discovered \emph{measurement induced phase transition} in monitored quantum systems, where the entanglement dynamics and purity of extensive quantum systems was found to undergo a transition in the long time limit as the measurement rate increases, see Refs.~\cite{skinner2019measurement,li2019measurement} and the review Ref.~\cite{potter2022entanglement}. While there is as of yet no formal connection between the phenomena explored here and those in the quantum setting, the resemblance between the two situations is suggestive.

\begin{figure}[h]
    \centering
    \includegraphics[width=\linewidth]{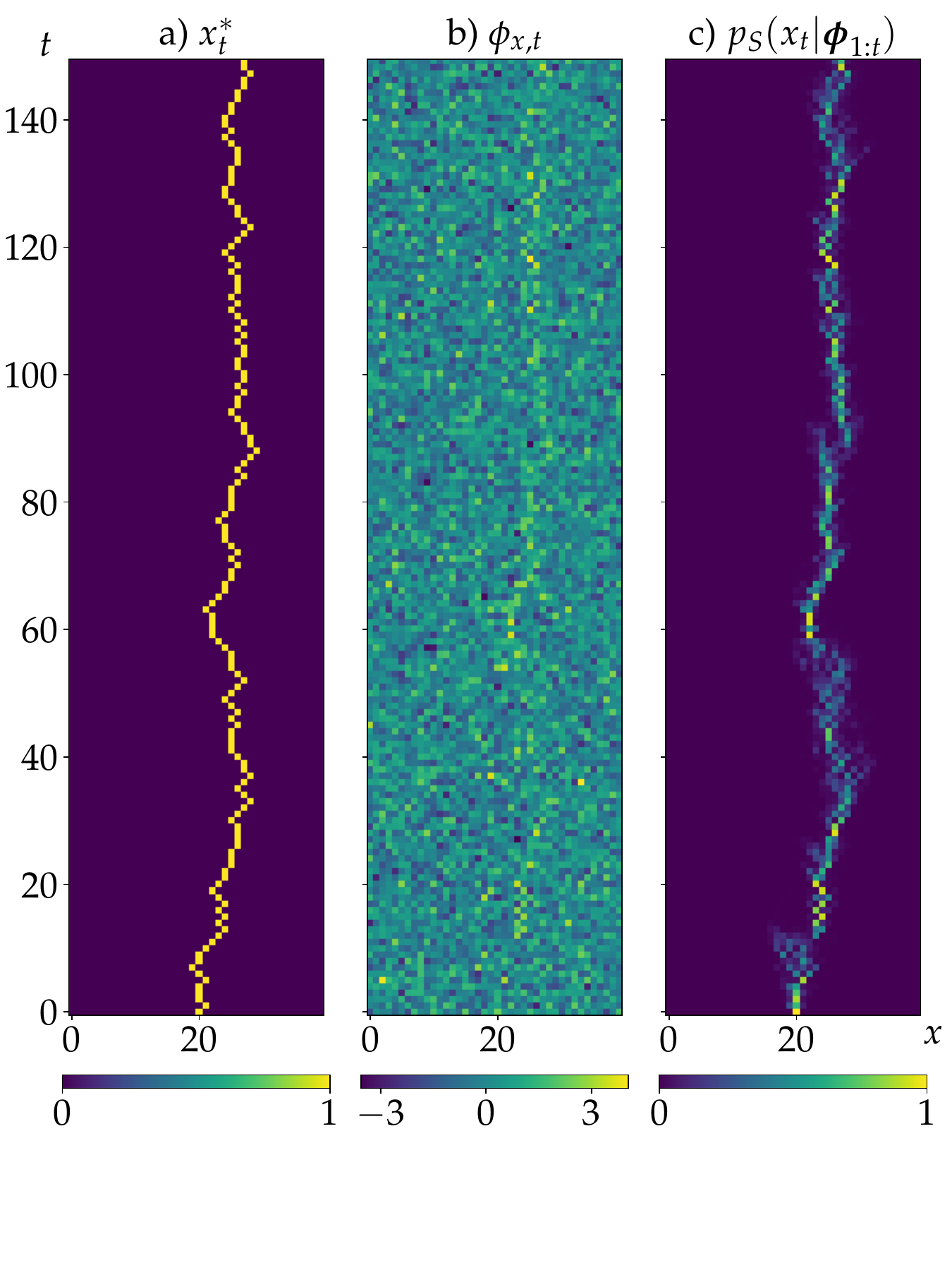}
    \caption{An example of Bayesian inference on a random walker from images with signal strength $\epsilon_\mathrm{T}=1.5$ and all other parameters $\sigma_\mathrm{T}=\sigma_\mathrm{S}=\epsilon_\mathrm{S}=1$. a) is a realisation of a random walker undergoing Brownian motion. b) shows the sets of `noisy images' taken at each timestep. c) is the posterior distribution inferred from the noisy images.}
    \label{fig:trajectories}
\end{figure}

The outline of the rest of this paper is as follows. In the next section we introduce the planted directed polymer problem and describe the relation to the directed polymer. In \cref{sec:1d} we study the one dimensional case numerically to establish some general features of the phenomenology of the problem. In \cref{sec:tree} we present analytical results for the case of paths on a tree, building on the approach of Ref.~\cite{derrida1988polymers} to the random polymer problem. Finally, we conclude in \cref{sec:conclusions} and offer a perspective on similar problems in this class.

\section{Statement of the planted directed polymer problem}

In this section we introduce the planted directed polymer problem. We describe the simplest case where the dynamics of our system is a random walk on the integers $\mathbb{Z}$: the generalisation \changes{of the model }to more complicated situations, including higher dimensions, should be obvious. The position of the walker at (integer) time $t$ is denoted by $x^*_t\in \mathbb{Z}$: throughout this work we follow Ref.~\cite{zdeborova2016statistical} in using $x_t^*$ to denote the `ground truth' or `planted' configuration, with  $x_t$ denoting the inferred configuration. The dynamics are specified by some transition kernel $p_\mathrm{T}(x^*_{t+1} \vert x^*_t)$, with initial distribution $p_\mathrm{T}
(x^*_1)$ (the meaning of the subscript T will be explained shortly). As our main example we will take a symmetric random walk (see \cref{fig:trajectories}a)
\begin{equation}
p_\mathrm{T}(x^*_{t+1} \vert x^*_t) = \frac{1}{2}\delta_{x^*_{t+1}, x^*_t} + \frac{1}{4}\delta_{x^*_{t+1}, x^*_t \pm 1}.
\end{equation}
Our goal is to infer the walker's trajectory from measurements that form noisy `images' specified by a `pixel' value $\phi_{x,t}$ at each position and time. These values are given by
\begin{align}\label{eq:measurement_model}
    \phi_{x, t} = \psi_{x, t} + \epsilon_\mathrm{T} \delta_{x, x^*_t},
\end{align}
where the pixel noise is a set of independent and identically distributed (iid) Gaussian distributions with standard deviation $\sigma_\mathrm{T}$, and $\epsilon_\mathrm{T}$ is the signal strength. An example is shown in \cref{fig:trajectories}b. We will sometimes use the standard probabilists' notation $\mathcal{N}(\mu,\sigma^2)$ for a normal distribution of mean $\mu$ and standard deviation $\sigma$, so that $\phi_{x, t}\sim p(\cdot |x^*_{t})=\mathcal{N}(\epsilon_\mathrm{T}\delta_{x_t,x^*_t}, \sigma_\mathrm{T}^2)$. We use the bold symbol $\bm{\phi}_t:=(\phi_{x,t}| x\in \mathbb{Z})$ for the entire image at some time $t$. Following \cite{zdeborova2016statistical}, we describe this process as the `teacher', and indicate its parameters using the subscript T.

Given $\phi_{x,t}$, the `student' can perform Bayesian inference to obtain the posterior over the walker's trajectory $x_{1:t}$ given the past history of noisy images, $\bm{\phi}_{1:t} := (\bm{\phi}_{\cdot, 1}, \dots, \bm{\phi}_{\cdot, t})$. We introduce the notation $X:=x_{1:t}$, $X^*:=x^*_{1:t}$ for the entire inferred and planted configurations respectively, and $\Phi:=\bm{\phi}_{1:t}$ for the sequence of images. The joint distribution is then
\begin{align} \label{eq:joint_path}
    p(X, \Phi, X^*) = p_\mathrm{S}(X \vert \Phi) p_\mathrm{T}(\Phi \vert X^*) p_\mathrm{T}(X^*).
\end{align}
$p_\mathrm{T}(\Phi \vert X^*)$ is the probability of obtaining the sequence of images conditioned on the path $X^*$, given by the above Gaussian distribution, and the student's posterior over the path $X = x_{1:t}$ is given by Bayes' rule
\begin{align} \label{eq:path_posterior}
    p_\mathrm{S}(X \vert \Phi) = \frac{p_\mathrm{S}(\Phi \vert X) p_\mathrm{S}(X)}{\sum_{X'} p_\mathrm{S}(\Phi \vert X') p_\mathrm{S}(X')}.
\end{align}
We use the subscript \text{S} to denote the student's distribution, which will allow us to consider a non Bayes-optimal setting in which the parameters of the student's model do not coincide with those of the teacher. Later, we will also consider the case where the student also does not have access to the functional form of the teacher model. Eq.~\eqref{eq:joint_path} defines the \emph{planted ensemble} (in the terminology of Ref.~\cite{zdeborova2016statistical}) for our problem.

Using the rules of conditional probability on the student's prior we can also write the joint probability as
\begin{align}
    p(X, \Phi, X^*) = \frac{p_\mathrm{S}(\Phi \vert X) p_\mathrm{S}(X) p_\mathrm{T}(\Phi \vert X^*) p_\mathrm{T}(X^*)}{p_\mathrm{S}(\Phi)},
\end{align}
where $p_\mathrm{S}(\Phi) = \sum_X p_\mathrm{S}(\Phi \vert X) p_\mathrm{S}(X)$. We see that in the Bayes optimal case, which is when the student and teacher distributions coincide, the joint distribution is symmetric with respect to $X \leftrightarrow X^*$. This means that inferred paths $X$ must have the same distribution as that of the true paths $X^*$. \changes{In the statistical physics of inference literature, this is also known as the Nishimori condition.}

\Cref{eq:joint_path} represents a complete description of the problem, but dealing with the distribution over the entire trajectory is not normally practical. Instead, it is more usual to focus on the posterior distribution of the present value given the measurement history i.e. $p_\mathrm{S}(x_t|\bm{\phi}_{1:t})$ (the \emph{filtering} problem). Below we present a tractable algorithm to calculate the posterior for the present value.

At $t=1$ Bayes' rule gives
\begin{align} \label{eq:initial_bayes}
    p_\mathrm{S}(x_1 \vert \bm{\phi}_{1}) = \frac{p_\mathrm{S}(\bm{\phi}_{1} \vert x_1) p_\mathrm{S}(x_1)}{\sum_{x'_1} p_\mathrm{S}(\bm{\phi}_{1} \vert x'_1) p_\mathrm{S}(x'_1)}.
\end{align}
$p_\mathrm{S}(x_1)$ is the prior distribution on the initial position. $p_\mathrm{S}(\bm{\phi}_1 \vert x_1)$ is the Gaussian distribution that follows from \cref{eq:measurement_model}. At subsequent times, the student uses the posterior of the previous timesteps to create the prior for the current timestep:
\begin{align} \label{eq:kushner}
    p_\mathrm{S}(x_t \vert \bm{\phi}_{1:t}) = \frac{p_\mathrm{S}(\bm{\phi}_{t} \vert x_t) p_\mathrm{S}(x_t \vert \bm{\phi}_{1:t-1})}{\sum_{x'_t} p_\mathrm{S}(\bm{\phi}_{t} \vert x'_t) p_\mathrm{S}(x'_t \vert \bm{\phi}_{1:t-1})},
\end{align}
where $p_\mathrm{S}(x_t \vert \bm{\phi}_{1:t-1})$ is given by the forward algorithm \cite{stratonovich1965conditional}
\begin{align} \label{eq:forward}
    p_\mathrm{S}(x_t \vert \bm{\phi}_{1:t-1}) = \sum_{x_{t-1}} p_\mathrm{S}(x_t \vert x_{t-1}) p_\mathrm{S}(x_{t-1} \vert \bm{\phi}_{1:t-1}),
\end{align}
and $p_\mathrm{S}(x_t \vert x_{t-1})$ is the transition kernel assumed by the student. An example of the posterior obtained in this way is depicted in \cref{fig:trajectories}c.

Alternatively, it is often convenient to deal with an \emph{unnormalised} posterior $q_\mathrm{S}(x_t|\phi_{1:t})$ that obeys the linear equation
\begin{align} \label{eq:zakai}
    q_\mathrm{S}(x_t \vert \bm{\phi}_{1:t}) = q_\mathrm{S}(\bm{\phi}_{t} \vert x_t) \sum_{x_{t-1}} q_\mathrm{S}(x_t \vert x_{t-1}) q_\mathrm{S}(x_{t-1} \vert \bm{\phi}_{1:t-1}),
\end{align}
involving the unnormalised analogues of the transition and measurement probabilities. The right hand side of \Cref{eq:path_posterior} and \Cref{eq:kushner} are both valid in terms of these unnormalised quantities, ex. in \cref{eq:kushner} the likelihood $p_\mathrm{S}(\phi_t \vert x_t)$ can be replaced with an unnormalised $q_\mathrm{S}(\phi_t \vert x_t)$. 



We are interested in distinguishing those regions in parameter space where the inferred trajectory $x_t$ closely tracks the planted configuration $x^*_t$ from those where it does not. We therefore consider the joint distribution
\begin{align} \label{eq:joint_phi}
    p(x_t, \bm{\phi}_{1:t}, x^*_{1:t}) = p_\mathrm{S}(x_t \vert \bm{\phi}_{1:t}) p_\mathrm{T}(\bm{\phi}_{1:t} \vert x^*_{1:t}) p_\mathrm{T}(x^*_{1:t}).
\end{align}

We have already discussed the first factor in \cref{eq:kushner}. The second factor corresponds to our Gaussian measurement model \cref{eq:measurement_model}. Finally, the third factor is the probability of a trajectory from the teacher Markov process:
\begin{align} \label{eq:planted-markov}
    p_\mathrm{T}(x^*_{1:t}) = \left(\prod_{\tau=2}^t p_\mathrm{T}(x^*_\tau \vert x^*_{\tau-1})\right) p_\mathrm{T}(x^*_1).
\end{align}

We can quantify the success of inference via the root mean squared error (RMSE) between the inferred position and true position at time $t$
\begin{align} \label{eq:mse}
    \text{RMSE}(t) := \sqrt{\E_{x_t, x^*_t}\left[(x_t - x^*_t)^2\right]},
\end{align}
which generalizes in an obvious way to the case of $d$-dimensions. Alternatively, we may consider the path-averaged quantity 
\begin{align}
    \overline{\text{RMSE}}(t) = \sqrt{\frac{1}{t} \E_{X, X^*}\left[ \sum_{\tau=1}^t (x_\tau - x^*_\tau)^2 \right]}.
\end{align}
If $\text{RMSE}(t)$ saturates at long times, $\text{RMSE}(\infty)$ and $\overline{\text{RMSE}}(\infty)$ will coincide.

In discrete settings we can also consider the average \emph{overlap}: the fraction of the time that the inferred path coincides with the true path
\begin{equation}
\overline{Y} := \frac{1}{t} \E_{X, X^*}\left[ \sum_{\tau=1}^t \delta_{x_\tau, x^*_\tau} \right].
\end{equation}
$\overline{Y}=1$ corresponds to perfect inference and $\overline{Y}\to 0$ at long times corresponds to failure of inference.

\subsection{Relation to the directed polymer problem}

Our Gaussian measurement model \cref{eq:measurement_model} gives the likelihood of an image $\bm{\phi}_t$ given current position $x_t$ (for either student S or teacher T variables)
\begin{align} \label{eq:likelihood}
    p(\bm{\phi}_t \vert x_t) & = \prod_{x'_t} \frac{1}{\sqrt{2 \pi \sigma^2}} \exp \left[ \frac{-(\phi_{x'_t, t} - \epsilon \delta_{x_t, x'_t})^2}{2 \sigma^2} \right] \\ 
    & = \exp\left(\left[\epsilon \phi_{x_t,t} - \epsilon^2/2\right]/\sigma^2\right)\pi(\bm{\phi}_{t}),
\end{align}
where $\pi(\bm{\phi}_{t})$ is the Gaussian measure with standard deviation $\sigma$ over all spatial points. Since the image at a given time is only conditionally dependent on position of the walker at that time, the likelihood for all images factorises as $p(\Phi \vert X) = \prod_{\tau=1}^t p(\bm{\phi}_\tau \vert x_\tau)$. As $x_t$ only enters in the  Boltzmann-like factor, \changes{and recalling that} the posterior \cref{eq:path_posterior} for the path $X = x_{1:t}$ can be written as \changes{a function of unnormalised prior $q(x_{1:t})$ and unnormalised likelihood,}
\begin{align} \label{eq:posterior}
    p(X \vert \Phi) = \frac{\exp\left[\frac{\epsilon}{\sigma^2} \sum_{\tau=1}^t \phi_{x_{\tau}, \tau} \right] q(x_{1:t})}{Z(\Phi)}.
\end{align}
The normalising factor, or partition function, is
\begin{align} \label{eq:partition_function}
    Z(\Phi) := \sum_{X} \exp\left[\frac{\epsilon}{\sigma^2} \sum_{\tau=1}^t \phi_{x_{\tau}, \tau} \right] q(x_{1:t}).
\end{align}

The partition function can be rewritten in a form that is more reminiscent of the directed polymer in a random medium,
\begin{align}
    Z(\Phi) = \sum_{X} \exp\left[ \sum_{\tau=1}^t h(x_{\tau-1:\tau}, \phi_\tau) \right] q(x_1),
\end{align}
with $h(x_{\tau-1:\tau}, \phi_\tau) := \ln q(x_\tau \vert x_{\tau-1}) + \frac{\epsilon}{\sigma^2} \phi_{x_\tau, \tau}$, corresponding to the `energy' of the polymer. The first term then corresponds to elasticity and the second a potential. Here we set $q(x_1 \vert x_0) = 1$ for convenience of notation.

Note that we can marginalise over the true paths $X^*$ in the joint distributions \cref{eq:joint_phi,eq:joint_path} and instead consider $p(X, \Phi) = p_\mathrm{S}(X \vert \Phi) p_\mathrm{T}(\Phi)$, where the `planted' disorder is distributed as
\begin{align} \label{eq:planted_disorder}
    p_\mathrm{T}(\Phi) & = \sum_{X^*} p_\mathrm{T}(\Phi \vert X^*) p_\mathrm{T}(X^*) \nonumber \\
    & \propto \pi_\mathrm{T}(\Phi) Z_\mathrm{T}(\Phi),
\end{align}
where $\pi_\mathrm{T}(\Phi)=\prod_t \pi_\mathrm{T}(\bm{\phi}_t)$. \cref{eq:planted_disorder} shows how the Gaussian measure $\pi_\mathrm{T}(\Phi)$ is modified by the presence of the ensemble of planted trajectories. When $\epsilon_\mathrm{T}=0$, $p_\mathrm{T}(\Phi)=\pi_\mathrm{T}(\Phi)$. In this case the joint probability $p(X, \Phi)=p_\mathrm{S}(X \vert \Phi)\pi_\mathrm{T}(\Phi)$ corresponds to that of the directed polymer in a random medium \cite{kardar2007statistical}. In terms of our inference problem, $\epsilon_\mathrm{T}=0$ means that the teacher provides no signal to the student, though the student believes a signal is present and uses $\epsilon_\mathrm{S}\neq 0$ to construct the posterior $p_\mathrm{S}(X \vert \Phi)$. As a result the path trajectory $x_{1:t}$ may be `pinned' in a random configuration by the `potential' $\phi_{x,t}$. When $\epsilon_\mathrm{T}\neq 0$ the teacher provides a signal to the student (c.f. \cref{fig:trajectories}) which alters the character of the posterior, as we will show in the next section.


Another way to view the problem is to separate $\Phi$ into random noise $\Psi := \bm{\psi}_{1:t}$ and signal $\epsilon_\mathrm{T}$. We will see that this form is more helpful for theoretical analysis. In this case the posterior becomes conditional on both $\Psi$ and $X^*$, as
\begin{align} \label{eq:psi}
    p_\mathrm{S}(X \vert \Psi, X^*) = \frac{\exp\left[ \sum_{\tau=1}^t h_{\mathrm{P}}(x_{\tau-1:\tau}, \psi_\tau, x^*_\tau) \right] q_\mathrm{S}(x_1)}{Z_{\mathrm{P}}(\Psi, X^*)},
\end{align}
where $h_{\mathrm{P}}(x_{\tau-1:\tau}, \psi_\tau, x^*_\tau) := \ln q_\mathrm{S}(x_\tau \vert x_{\tau-1}) + \frac{\epsilon_\mathrm{S}}{\sigma_\mathrm{S}^2} \psi_{x_\tau, \tau} + \frac{\epsilon_\mathrm{S} \epsilon_\mathrm{T}}{\sigma_\mathrm{S}^2} \delta_{x_\tau, x^*_\tau} $ and the \textit{planted} partition function is
\begin{align}
    Z_{\mathrm{P}}(\Psi, X^*) := \sum_X \exp\left[ \sum_{\tau=1}^t h_{\mathrm{P}}(x_{\tau-1:\tau} \psi_\tau, x^*_\tau) \right] q_\mathrm{S}(x_1).
\end{align}
Then $p(\Psi) = \pi_\mathrm{T}(\Psi)$ is just the iid Gaussian measure with teacher's parameters.

\section{1d case}\label{sec:1d}

In this Section we present the results of numerical investigations of the properties of the posterior $p_\mathrm{S}(X|\Phi)$ for a one-dimensional walker, and then further develop the above theoretical picture.

\subsection{Numerical investigation}\label{sec:numerics}

\begin{figure}
    \centering
    \includegraphics[width=\linewidth]{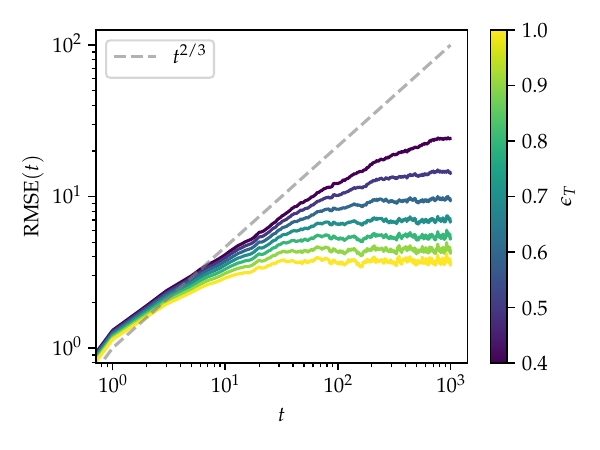}
    \caption{Plot of root-mean-squared error between true and inferred paths over time $t$ for $\epsilon_\mathrm{S}=\sigma_\mathrm{T}=\sigma_\mathrm{S}=1$. As teacher signal strength $\epsilon_\mathrm{T}$ increases, the root-mean-squared error plateaus at smaller values.}
    \label{fig:rms_dist_plateau} 
\end{figure}

\begin{figure}
    \centering
    \includegraphics[width=0.7\linewidth]{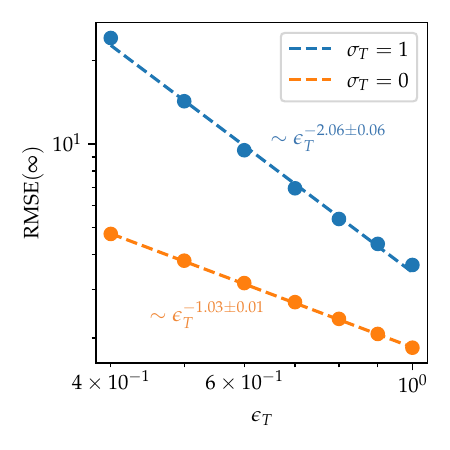}
    \caption{Scaling of estimated long time $\text{RMSE}(\infty)$ with teacher signal strength $\epsilon_\mathrm{T}$ for teacher noise strength $\sigma_\mathrm{T}=0$ and $\sigma_\mathrm{T} = 1$. Both cases support power law with $\epsilon_\mathrm{T}$, indicating a finite RMSE with any signal. The numerical results for $\sigma_\mathrm{T}=0$ case confirms the $\propto \epsilon_\mathrm{T}^{-1}$ theoretical prediction, \cref{eq:rms_dotsenko}. The $\sigma_\mathrm{T}=1$ is also consistent with $\propto \epsilon_\mathrm{T}^{-2}$ from our quantitative argument, \cref{eq:x_plateau}.}
    \label{fig:rms_dist_scaling}
\end{figure}

\begin{figure}
    \center
    \begin{tikzpicture}[scale=1.5, samples=100]
        \node at (0,0) (nodeA) {};
        \node at (3,3) (nodeB) {};
        \draw[->, color=red] (0, 0) -- (3, 0) node[right] {$\epsilon_\mathrm{S}$} node[midway, below, sloped] {$\alpha=\frac{2}{3}$};
        \draw[->, color=blue] (0, 0) -- (0, 3) node[above] {$\epsilon_\mathrm{T}$} node[midway, above, sloped] {$\alpha=\frac{1}{2}$};
        \draw[-, color=teal] (2, 1) node[below] {$\alpha=\frac{1}{2}$?};
        \draw[-] (0, 0) node[below left] {$0$};
        \draw[dashed, color=blue] (0, 0) -- (2.75, 2.75) node [right, sloped] (TextNode) {$\epsilon_\mathrm{S}=\epsilon_\mathrm{T}$} node [midway, below, sloped] (TextNode) {$\alpha=\frac{1}{2}$};
    \end{tikzpicture}
    \caption{Conjectured phase diagram for wandering exponents $\alpha$ in the $\epsilon_\mathrm{S}-\epsilon_\mathrm{T}$ plane with fixed $\sigma_\mathrm{S}=\sigma_\mathrm{T}=\sigma$. The blue and red texts are analytical results. Green is the conjectured behaviour throughout the plane excluding $\epsilon_\mathrm{T}=0$. Numerics appears to support that $\alpha=1/2$ for any finite $\epsilon_\mathrm{T}>0$.}
    \label{fig:wandering_Schematic}
\end{figure}
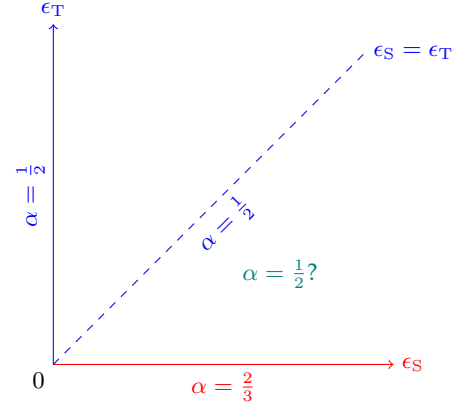

We solve \cref{eq:kushner} numerically using $\bm{\phi}_t$ generated using $p_\mathrm{T}(\bm{\phi}_t|x^*_t)$ with $x^*_t$ sampled from a symmetric random walk. We set $\epsilon_\mathrm{S}=\sigma_\mathrm{T}=\sigma_\mathrm{S}=1$ unless otherwise specified. \Cref{fig:rms_dist_plateau} shows the RMSE \cref{eq:mse} between the true and inferred positions, $\text{RMSE}(t)$ with maximum time $t_{\max} = 1000$, system size $L=1000$ with periodic boundary conditions, as $\epsilon_\mathrm{T}$ is varied. We see that for $\epsilon_\mathrm{T}$ close to $1$, $\text{RMSE}(t)$ appears to plateau within $t_\mathrm{plateau}$ at some $\text{RMSE}(\infty)$. In \cref{fig:rms_dist_scaling}, for the values of $\epsilon_\mathrm{T}$ where we can see a clear plateau, we study the scaling of estimated long-time $\text{RMSE}(\infty)$ with $\epsilon_\mathrm{T}$. For $\sigma_\mathrm{T} = 1$ We see good evidence supporting that $\text{RMSE}(\infty) \sim \epsilon_\mathrm{T}^{-2.06(1)}$, which, if the scaling holds for all values of $\epsilon_\mathrm{T} > 0$, suggests that $\text{RMSE}(t)$ is bounded for any $\epsilon_\mathrm{T} > 0$. This suggests that $\alpha=1/2$ for any finite $\epsilon_\mathrm{T} > 0$. For the $\sigma_\mathrm{T} = 0$ case, the numerics support the $\text{RMSE}(\infty) \sim \epsilon_\mathrm{T}^{-1}$ scaling of Eq. \eqref{eq:rms_dotsenko}. This suggests that there is a change in the scaling of $\text{RMSE}(\infty)$ with and without the presence of disorder. We summarise our discussion on the wandering exponents in a conjectured phase diagram in \cref{fig:wandering_Schematic}.

In the directed polymer problem $\epsilon_\mathrm{T}=0$, there is no transition at finite noise strength $\sigma_\mathrm{T}$, and the wandering exponent is $\alpha = 2/3$ for any $\sigma_\mathrm{T} > 0$. Our numerics seem to suggest that the wandering exponent switches immediately to $\alpha=1/2$ for any finite $\epsilon_\mathrm{T} > 0$. In the language of directed polymers, $\epsilon_\mathrm{T} > 0$ corresponds to the strength of an attractive delta potential. Our numerics suggest a hierarchy of terms in the polymer's `energy' \cref{eq:energy}: The disordered media term always overcomes the polymer `elasticity' term, while the delta-potential term always beats the disordered media term. In the language of Bayesian inference, inference is always possible with any true signal strength.


\begin{figure}[H]
    \centering
    \includegraphics[width=\linewidth]{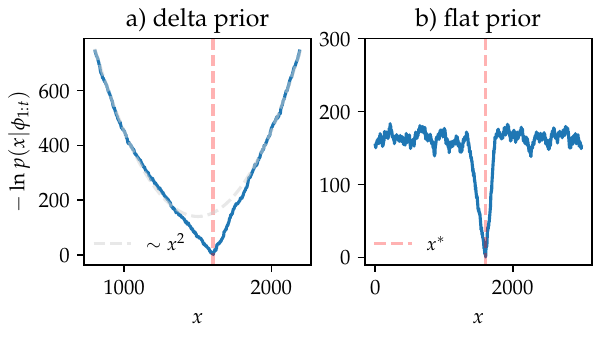}
    \caption{An example of the free energy profile $F = -\ln p_\mathrm{S}(x_t \vert \phi_{1:t})$ at $t=1000$ for a) delta peak prior b) flat prior. Vertical dashed pink line indicates true position $x^*$ and dashed grey line indicates Gaussian modulation due to the delta peak prior.}
    \label{fig:free_energy_profiles}
\end{figure}

\subsection{Directed polymer mapping} \label{sec:dp-mapping}

Here, to help with theoretical analysis, we take the continuum limit. Let us work with the separated noise and signal picture as Eq. \eqref{eq:psi}.  We define $\tilde x = a x$, $\tilde \tau = a \tau$, with final time $\tilde{t} = a t$, then take the $a \rightarrow 0$ limit with $\tilde{t}$ fixed. Let us assume that the transition kernel $q_\mathrm{S}(x_t \vert x_{t-1})$ only depends on the distance between the two points, and that it is maximised when the distance is zero. We take the scaling form $\ln q_\mathrm{S} (x_t \vert x_{t-1}) = f_\mathrm{S}(\sqrt{a}[x_t - x_{t-1}])$ then expand around $0$. We write $\psi_{x, t} = \sigma_\mathrm{T} \eta_{x,t}$, where $\eta_{x, t} \sim \mathcal{N}(0, 1)$. Hence we can write the unnormalised posterior as
\begin{widetext}
\begin{align}
    q_{\mathrm{S}}(x_{t} \vert \bm{\psi}_{1:t}, x^*_{1:t}) = \sum_{x_{t-1} \cdots x_1} \exp\left(\sum_{\tau=1}^{t} \left[-\frac{a}{4 \nu_\mathrm{S}} \left[x_\tau - x_{\tau-1}\right]^2 + \frac{\epsilon_\mathrm{S} \sigma_\mathrm{T}}{\sigma_\mathrm{S}^2} \eta_{x_\tau, \tau} + \frac{\epsilon_\mathrm{S} \epsilon_\mathrm{T} }{\sigma_\mathrm{S}^2} \delta_{x_\tau, x^*_\tau} \right] + \mathcal{O}(a^{3/2})\right) q_{\mathrm{S}}(x_1),
\end{align}
where $\nu_\mathrm{S} = \frac{1}{2 \lvert f''_\mathrm{S}(0) \rvert}$ is the student's width of the kernel, and $x_0 = x_1$ for notational convenience. In the $a\rightarrow 0$ limit we have $\sum_\tau a \rightarrow \int d\tau$, $\frac{1}{a} \delta_{x_t, x^*_t} \rightarrow \delta(\tilde{x}(\tilde{\tau}) - \tilde{x}^*(\tilde{\tau}))$, and $\frac{ax_{\tau} - ax_{\tau-1}}{a} \rightarrow \frac{d \tilde{x}}{ d \tilde{t}}$. We define $\tilde{\eta}_{x, t} := \frac{1}{a} \eta_{x,t}$ to retrieve white noise at continuum, $\mathbb{E}_{\tilde{\eta}} \tilde{\eta}_{x, t} \tilde{\eta}_{x', t'} = \frac{1}{a} \delta_{x, x'} \frac{1}{a}\delta_{t, t'} \rightarrow \delta(\tilde{x}-\tilde{x}') \delta(\tilde{t} - \tilde{t}')$. Taking the limit then relabeling the continuum variables by removing the tildes, we have
\begin{align}
    q_{\mathrm{S}}(x_t \vert \eta, x^*) \sim \int_{p_\mathrm{S}(x(0))}^{x(t)=x_t} D[x] e^{- E[x, \eta, x^*]},
\end{align}
where now $\eta$ and $x^*$ refer to functions or continuous paths, and the `energy' of the polymer is
\begin{align} \label{eq:energy}
    E[x, \eta, x^*] = \int_0^t d\tau\left[ \frac{1}{4 \nu_\mathrm{S}} \dot{x}^2 - \frac{\epsilon_\mathrm{S} \sigma_\mathrm{T}}{\sigma_\mathrm{S}^2} \eta(x(\tau), \tau) - \frac{\epsilon_\mathrm{S} \epsilon_\mathrm{T}}{\sigma_\mathrm{S}^2} \delta(x(\tau) - x^*(\tau))\right].
\end{align}
\end{widetext}
This is the directed polymer with an attractive moving delta potential located at the path of the random walker $x^*(t)$. $p_\mathrm{S}(x(0))$ indicates the initial prior distribution. For example, for the Kronecker-delta initial state $p_\mathrm{S}(x_1) = \delta_{x_1, 0}$ we can replace $p_\mathrm{S}(x(0)) \rightarrow x(0)=0$.

Taking the same limit for the partition function, the normalised posterior can then be written as
\begin{align}
    p_{\mathrm{S}}(x_t \vert \eta, x^*) = \ddfrac{
        q_{\mathrm{S}}(x_t \vert \eta, x^*)
    }{
        Z_{\mathrm{P}}(t)
    },
\end{align}
where $Z_{\mathrm{P}}(t) = \int_{p_\mathrm{S}(x(0))} D[x] e^{-E[x,\eta, x^*]}$. In the continuum limit, we have $p_\mathrm{T}(x_{1:t}^*) \sim e^{-E_0[x^*]} p_\mathrm{T}(x^*(0))$ where $E_0 = \int_0^t d \tau \frac{1}{4 \nu_\mathrm{T}}{(\dot{x}^*)}^2$. Marginalising over $\bm{\eta}_{1:t}$ and $x^*_{1:t-1}$ then taking the continuum limit all together, the joint distribution for inferred position and true position at time $t$ is given by
\begin{equation}
    \begin{aligned}
        p(x_t, x^*_t) & = \int D \pi[\eta] \int_{p_\mathrm{T}(x^*(0))}^{x^*(t)=x^*_t} D[x^*] e^{-E_0[x^*]} 
    \frac{
        q_{\mathrm{S}}(x_t \vert \eta, x^*)
    }{
        Z_{\mathrm{P}}(t)
    }.
    \end{aligned}
\end{equation}
Just as the directed polymer can be mapped to the KPZ equation and a diffusion process with random sources and sinks, from the path integral picture, we can see that the unnormalised posterior, with abuse of notation $q_{\mathrm{S}}(x, t) = q_{\mathrm{S}}(x_t \vert \eta, x^*)$, should evolve as a random diffusion process with a delta potential,
\begin{align} \label{eq:schrodinger}
    \frac{\partial q_{\mathrm{S}}}{\partial t} = \nu_\mathrm{S} \partial_x^2 q_{\mathrm{S}} + \left( \frac{\epsilon_\mathrm{S} \sigma_\mathrm{T}}{\sigma_\mathrm{S}^2} \eta(x,t) + \frac{\epsilon_\mathrm{S} \epsilon_\mathrm{T}}{\sigma_\mathrm{S}^2} \delta(x - x^*(t))\right) q_{\mathrm{S}},
\end{align}
and that the `free energy' \changes{$F := -\ln q_{\mathrm{S}}(x,t)$} evolves as a KPZ equation with a delta potential,
\changes{
\begin{align} \label{eq:kpz}
    \frac{\partial F}{\partial t} & = \nu_\mathrm{S} \partial_x^2 F - \nu_\mathrm{S} (\partial_x F)^2 \nonumber \\ & - \frac{\epsilon_\mathrm{S} \sigma_\mathrm{T}}{\sigma_\mathrm{S}^2} \eta(x,t) - \frac{\epsilon_\mathrm{S} \epsilon_\mathrm{T}}{\sigma_\mathrm{S}^2} \delta(x - x^*(t)).
\end{align}
}

{In \cref{fig:free_energy_profiles}, we plot a typical free energy profile for a given random walk $x^*$ starting from both flat and delta initial priors $p_\mathrm{S}(x_1) \sim 1$ and $\delta_{x_t, 0}$ respectively. We see that, \changes{superimposed on} the rough landscape corresponding to KPZ (Kardar-Parisi-Zhang) behaviour, \changes{there is} a \changes{linear `wedge'} in the free energy. In the noiseless case, $\sigma_\mathrm{T} = 0$, Eq. \eqref{eq:schrodinger} is the imaginary-time Schrödinger equation for a time-varying Hamiltonian with a delta potential located at $x^*$, $\hat{H}(t) = - \nu_\mathrm{S} \partial^2_x - \frac{\epsilon_\mathrm{S} \epsilon_\mathrm{T}}{\sigma_\mathrm{S}^2} \delta(x-x^*(t))$. The instantaneous ground state of $\hat H(t)$ is the single bound state given by $q_0(x) \sim e^{- \kappa \lvert x - x^* \rvert}$, with $\kappa = \frac{\epsilon_\mathrm{S} \epsilon_\mathrm{T}}{2 \nu_\mathrm{S} \sigma_\mathrm{S}^2}$ and energy $E_0=-\nu_\mathrm{S}\kappa^2$. The imaginary time evolution projects \changes{onto} the ground state, and if the potential moves slowly in time, we expect the $q_{\mathrm{S}}(x, t)$ to follow the potential, so that $q_{\mathrm{S}}(x,t) \sim e^{-\kappa \lvert x - x^*(t) \rvert + \nu_\mathrm{S} \kappa^2 t}$. Taking the logarithm, the free energy should have the form \changes{$F_\mathrm{gs} \sim \kappa \lvert x - x^* \rvert - \nu_\mathrm{S} \kappa^2 t$}.}

\usetikzlibrary{decorations.pathmorphing}

\pgfdeclaredecoration{jiggly}{step}
{
  \state{step}[width=+\pgfdecorationsegmentlength]
  {\pgfmathsetmacro{\delta}{rand*\pgfdecorationsegmentamplitude}
    \pgfpathlineto{
      \pgfpointadd
      {\pgfpoint{\pgfdecorationsegmentlength}{0pt}}
      {\pgfpointpolar{90-\pgfdecoratedangle}
          {\delta}}
    }
  }
  \state{final}
  {
    \pgfpathlineto{\pgfpointdecoratedpathlast}
  }
}
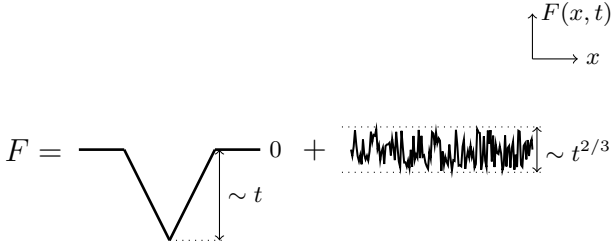
\begin{figure}[h]
    \center 
    \begin{tikzpicture}[scale=0.6, samples=100]
        \node at (-6, 2) {$\;$};
        \node at (-6, 0) {\large $F =$};
        \draw[->] (5, 2) -- (6, 2) node[right] {\small $x$};
        \draw[->] (5, 2) -- (5, 3) node[right] {\small $F(x, t)$};
        \begin{scope}[shift={(-3, 0)}]
            \draw[line width=1] (-2, 0) -- (-1, 0);
            \draw[line width=1] (-1, 0) -- (0, -2);
            \draw[line width=1] (0, -2) -- (1, 0);
            \draw[line width=1] (1, 0) -- (2, 0) node[right] {$0$};
            \draw[dotted] (0, -2) -- (1.2, -2);
            \draw[<->] (1.1, 0) -- node[right]{$\sim t$} (1.1, -2);
        \end{scope}

        \node at (0.2, 0) {\large$+$};

        \pgfmathsetseed{1}
        \begin{scope}[shift={(3, 0)}]
            \draw[line width=0.75] {decorate[decoration={jiggly, segment length=0.5,amplitude=8}]{(-2,0) -- (2,0)}};
            \draw[dotted] (-2.2, 0.5) -- (2.2, 0.5);
            \draw[dotted] (-2.2, -0.5) -- (2.2, -0.5);
            \draw[<->] (2.1, -0.5) -- node[right]{$\sim t^{2/3}$} (2.1, 0.5);
        \end{scope}
        \node at (6, 2) {$\;$};

    \end{tikzpicture}
    \caption{Schematic of the competition between the free energy of the planting $F_\mathrm{plant} \sim t$ (left term) vs. the KPZ free energy $F_\mathrm{KPZ} \sim t^{2/3}$ (right term). At early times $F_\mathrm{KPZ}$ beats $F_\mathrm{plant}$ while eventually $F_\mathrm{plant}$ always wins.}
    \label{fig:plants-vs-zombies}
    \end{figure}

\changes{We now present a heuristic picture of how this behaviour is modified for $\sigma_\mathrm{T}>0$, which allows us to quantitiatively describe the saturation of the mean-squared error in \Cref{fig:rms_dist_plateau,fig:rms_dist_scaling}. We compare the free energy profiles for the $\sigma_\mathrm{T}=0$ and $\epsilon_\mathrm{T}=0$ cases. 

First, for $\sigma_\mathrm{T}=0$ and $\epsilon_\mathrm{T}>0$, $F_\mathrm{gs}$ discussed above cannot be the solution for early times, since if we were to start off with a flat initial condition, we would have $F(x, 0) = \mathrm{cst}$, but the long-time solution above has $F_\mathrm{gs}(\pm \infty, t) = \infty$. We ignore the motion of $x^*$, setting $x^*=0$, and will justify this assumption afterwards. To look at the early-time transient behavior, we first relate \cref{eq:kpz} to a Burgers-like equation for $u(x,t) = 2\nu_\mathrm{S} \partial_x F(x,t)$
\begin{equation} \label{eq:burgers}
    \partial_t u + u \partial_x u = \nu_\mathrm{S}\partial_x^2 u - 4\nu_\mathrm{S}^2\kappa \delta'(x).
\end{equation}
The Burgers equation (i.e. without the $\delta'(x)$ on the RHS) has right- and left-moving soliton solutions \cite{olver2014introduction} of the form $u_\mathrm{sol}(x-vt)$ and $-u_\mathrm{sol}(-x-vt)$, where 
\begin{equation} \label{eq:soliton}
u_\mathrm{sol}(z) = \frac{2v}{1+\exp\left[(v/\nu_\mathrm{S})z\right]}.
\end{equation}
We therefore make the ansatz, which we expect to be valid at long times
\begin{equation} \label{eq:u-planted}
    u_\mathrm{plant}(x,t) = \begin{cases}
        u_\mathrm{sol}(x-vt) & x\geq 0, \\
        -u_\mathrm{sol}(-x-vt) & x<0. 
    \end{cases}
\end{equation}
This ansatz has a jump at the origin that approaches $4v$ at long times. The $\delta'(x)$ term in \cref{eq:burgers} fixes $v=\nu_\mathrm{S} \kappa$.

The free energy is given by $F(x,t)=\frac{1}{2\nu_\mathrm{S}}\int^x u(x',t) dx'+ c(t)$, with the constant of integration $c(t)$ chosen so that $F(\pm \infty,t)=0$. In the limit that the width of the solitons can be neglected, $F(x,t)$ takes the form
\begin{align} \label{eq:ansatz}
    F_\mathrm{plant}(x,t) = 
    \begin{cases} 
        -\nu_\mathrm{S} \kappa^2 t + \kappa\abs{x} & \lvert x \rvert \leq \nu_\mathrm{S} \kappa t \\ 
        0 & \lvert x \rvert > \nu_\mathrm{S} \kappa t. 
    \end{cases}
\end{align}
This is a `wedge' shape that is minimised at $x=0$ that `carves' out downwards in time, shown on the left of \cref{fig:plants-vs-zombies}. Our approximation that $x^*=0$ is seen to be justified at long times because of the ballistic spreading of the wedge, which overwhelms the diffusive motion of $x^*$.


Next, we look at the known scaling of the strong coupling KPZ fixed point of the unplanted case $\epsilon_\mathrm{T} = 0$, $\sigma_\mathrm{T} > 0$. Ref. \cite{calabrese2011exact} finds the exact scaling of the `height field' $h(x, t) := - F(x, t)$ again for the flat initial condition. The height field evolves as $\partial_t h = \nu \partial_x^2 h +\frac{\lambda_0}{2} (\partial_x h)^2 + \sqrt{D} \eta$. From this we can read off the mapping  $\nu = \nu_\mathrm{S}$, $\lambda_0/2 = \nu_\mathrm{S}$, $\sqrt{D} = \epsilon_\mathrm{S} \sigma_\mathrm{T} / \sigma_\mathrm{S}^2$. They show that the height field has an asymptotic form $\frac{\lambda_0}{2 \nu} h_x = v_0 t +\lambda \xi_{x,t}$, where $\xi_{x, t} \sim \mathcal{O}(1)$ is a random field in space at each time, as sketched on the right of \cref{fig:plants-vs-zombies}, $\lambda = (1/2)(\bar{c}^2 t/T^5)^{1/3}$, $\bar{c} = D\lambda_0^2$ and $T=2 \nu$. $v_0$ is a constant background and therefore can be ignored. Converting from their variables to ours, we find that
\begin{align}
    F_\mathrm{KPZ}(x,t) = 2^{-4/3} \epsilon_\mathrm{S}^{4/3} \sigma_\mathrm{T}^{4/3} \sigma_\mathrm{S}^{-8/3} \nu_\mathrm{S}^{-1/3} t^{1/3} \xi_{x, t}.
\end{align}
We now interpret the saturation of the mean squared error as the result of competition between $F_\mathrm{plant}(x,t)$ and $F_\mathrm{KPZ}(x,t)$, as illustrated schematically in \cref{fig:plants-vs-zombies}. Due to the $t$ vs. $t^{2/3}$ scaling, we expect the KPZ effect to win until the planting effect is stronger, therefore when $F_\mathrm{plant}(0, t_\mathrm{plateau}) \sim F_\mathrm{KPZ}(0, t_\mathrm{plateau})$. This leads to the estimate
\begin{align}
    t_\mathrm{plateau} \sim \frac{\sigma_\mathrm{S}^2 \sigma_\mathrm{T}^2 \nu_\mathrm{S}}{\epsilon_\mathrm{S} \epsilon_\mathrm{T}^3}.
\end{align}
From \cref{fig:rms_dist_plateau}, we have the picture that the student's walker grows as $x \sim t^{2/3}$ until we hit $t_\mathrm{plateau}$, where it plateaus at $\overline{\mathrm{RMSE}}(\infty) = x_\mathrm{plateau} \sim t_\mathrm{plateau}^{2/3}$. Substituting this in, we find that
\begin{align} \label{eq:x_plateau}
    \overline{\mathrm{RMSE}}(\infty) \sim \frac{\sigma_\mathrm{S}^{4/3} \sigma_\mathrm{T}^{4/3} \nu_\mathrm{S}^{2/3}}{\epsilon_\mathrm{S}^{2/3} \epsilon_\mathrm{T}^{2}},
\end{align}
from which we predict that $x_\mathrm{plateau} \sim \epsilon_\mathrm{T}^{-2}$. This is consistent with the scalings we see in \cref{fig:rms_dist_scaling}.

The above picture supports an absence of a phase transition for 1D, as the scaling holds for any parameters as long as they are not zero. This is because for the unplanted case $\epsilon_\mathrm{T} = 0$, for any $\sigma_\mathrm{T} > 0$, the system is always in the strong coupling KPZ fixed point, which in turn is a consequence of the recurrence of random walks in 1D. This is not true in the 3D case, where there is a transition between the weak and strong disorder phases. Therefore, we may expect a transition in that case. We also note that the case of the delta initial condition would require a more involved ansatz but we expect the results to be similar due to the same underlying scaling.
}

In the planted ensemble picture, we can consider the posterior for path $x_{1:t}$ given $\bm{\phi}_{1:t}$. Following the same procedure to go to the continuum, observables $O[x, x^*]$ dependent on entire paths can be formally written as
\begin{equation}
    \begin{aligned}
        \mathop{\mathbb{E}}_{x, \eta, x^*} O[x, x^*] & = \int D\pi[\eta] \int_{p_\mathrm{T}(x^*(0))} D[x^*] e^{-E_0[x^*]} \\ & \times \frac{\int_{p_\mathrm{S}(x(0))} D[x] O[x, x^*] e^{-E[x, \eta, x^*]}}{Z_{\mathrm{P}}(t)}.
    \end{aligned}
\end{equation}

The case $\sigma_\mathrm{T} = 0$ with the delta initial condition is the short-range interaction limit for the `memory model' considered in Ref. \cite{dotsenko2022one}, but with open boundary condition at the end times. This case corresponds to the somewhat unrealistic scenario where the student does not know that the images given by the teacher have no noise, and assumes that there is some level of noise. Ref. \cite{dotsenko2022one} considers the long-time path-averaged mean squared distance between the student and teacher paths, $\overline{\mathrm{RMSE}}(\infty)$. Translating between the two languages, $1/4 \nu_\mathrm{S} = \beta/2$, $\beta u = \epsilon_\mathrm{S} \epsilon_\mathrm{T} / \sigma_\mathrm{S}^2$, and $1/4\nu_\mathrm{T} = \gamma/2$, we find
\begin{align} \label{eq:rms_dotsenko}
    \overline{\mathrm{RMSE}}_{\sigma_\mathrm{T} = 0}(\infty) \sim \ddfrac{\sigma_\mathrm{S}^2}{\epsilon_\mathrm{S} \epsilon_\mathrm{T}} (\nu_\mathrm{S} + \nu_\mathrm{T}).
\end{align}

For the delta initial condition, a quantity of interest is the long-time wandering exponent $\alpha$ that describes the long-time growth of fluctuations $\sqrt{\E[ x^2_t]} \sim t^\alpha$. In the case $\epsilon_\mathrm{T} = 0$, we have the directed polymer and have $\alpha=2/3$. In case $\epsilon_\mathrm{S} = 0$, looking at Eq. \eqref{eq:schrodinger}, we have the heat equation for $q_\mathrm{S}$ and have $\alpha=1/2$. In the Bayesian optimal case, $\mathrm{S}=\mathrm{T}$, since the joint distribution is symmetric with $x_{1:t}$ and $x^*_{1:t}$, and because the teacher $x^*_{1:t}$ undergoes a random walk, we must also have $\alpha=1/2$. What happens away from these limits is not immediately obvious. 

\begin{figure}
    \centering
    \includegraphics[width=\linewidth]{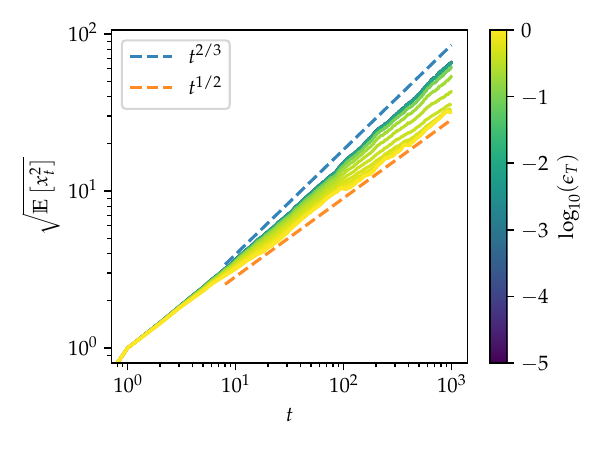}
    \caption{Plot of root-mean-squared width of the inferred path $\sqrt{\E[x^2_t]}$. At high teacher signal strength $\epsilon_\mathrm{T}$ we see that RMS width switches from $\sim t^{2/3}$ to $\sim t^{1/2}$ behaviour at plateau time. As $\epsilon_\mathrm{T}$ is lowered we see a crossover to $\sim t^{2/3}$ behaviour at finite time.}
    \label{fig:rms_crossover}
\end{figure}

Let us argue that $\alpha=1/2$ if $\overline{\mathrm{RMSE}}(\infty)$ is finite for a fixed $\epsilon_\mathrm{T}$. Since the RMS distance between the true and inferred paths is bounded, $x_t \sim x^*_t + C$ with $\lvert C \rvert \lesssim \overline{\mathrm{RMSE}}(\infty)$. Since $x^*_t$ can grow arbitrarily with time, at long times the $x^*_t$ term dominates and fluctuations of $x_t$ should grow as $x^*_t$ does. In \cref{fig:rms_crossover} we plot the RMS width of the inferred position, $\sqrt{\E[x^2_t]}$. We see that for large signal strength $\epsilon_\mathrm{T} \sim 1$, the RMS width switches from $\sim t^{2/3}$ to $\sim t^{1/2}$ at some time $t_\mathrm{plateau}$. Furthermore as $\epsilon_\mathrm{T}$ is lowered we see a crossover from $\alpha=1/2$ to $\alpha=2/3$ for a given finite $t$.

\section{Tree case} \label{sec:tree}

\begin{figure}[h]
    \centering
    \includegraphics[width=\linewidth]{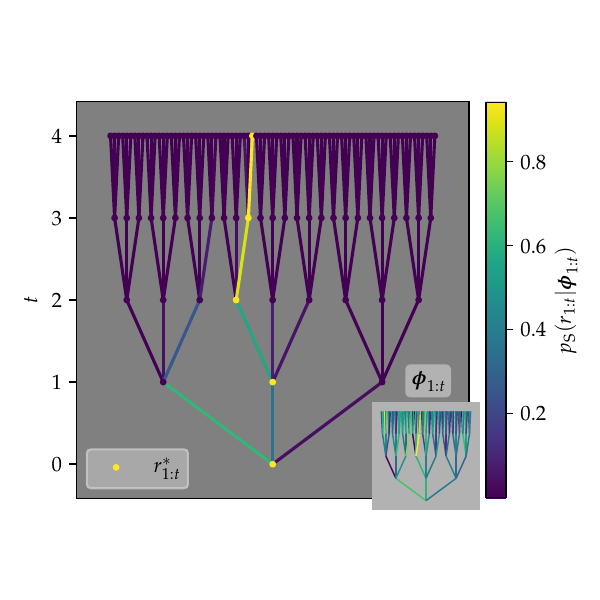}
    \caption{An example of Bayesian inference on a random walker on a tree with branching number $k=3$, $t=4$, teacher and student signal strengths $\epsilon_\mathrm{T}=\epsilon_{\mathrm{S}}=2$, and $\sigma_\mathrm{T}=\sigma_\mathrm{S}=1$. The yellow nodes in the main figure are the true positions of the random walker while the colour of each edge corresponds to the posterior for the path ending at the next node. (Inset) The `noisy images' taken at each timestep.}
    \label{fig:tree_trajectories}
\end{figure}

\changes{We now} consider the case where the random walker moves in a Cayley tree of depth $t$ with branching number $k$, moving deeper into the tree at each timestep, as shown by the location of the yellow nodes in \cref{fig:tree_trajectories}. \changes{The Cayley tree can be seen as a Bethe approximation to a lattice as spatial dimension goes to infinity \cite{katsura1974bethe}}. On the tree, positions can be labeled by which direction the walker took at each timestep. Therefore we use the string notation $r_{1:t} = r_1 r_2 \dots r_t$ to denote positions on the tree, where each $r_\tau = 1, \dots, k$ denotes direction chosen at depth $\tau$. On a tree, there is only one path for a given position. Therefore we also can interpret $r_{1:t}$ as the path taken. If we assume that all branches are equally likely, i.e.
\begin{align}
    p(r'_{1:t} \vert r_{1:t-1}) = 
    \begin{cases}
        \frac{1}{k} & \text{if }r'_{1:t-1} = r_{1:t-1}, \\
        0 & \text{otherwise},
    \end{cases}
\end{align}
then the posterior can be written as
\begin{align}
    p_\mathrm{S}(r_{1:t} \vert \phi_{1:t}) = \frac{e^{\epsilon_\mathrm{S} \Phi(r_{1:t})/\sigma_\mathrm{S}^2}}{Z_\mathrm{S}(\Phi)},
\end{align}
where we define $\Phi(r_{1:t}) := \phi_{r_1, 1} + \phi_{r_{1:2}, 2} + \cdots + \phi_{r_{1:t}, t}$, and $Z_\mathrm{S}(\Phi) = \sum_{r_{1:t}}e^{\epsilon_\mathrm{S} \Phi(r_{1:t})/\sigma_\mathrm{S}^2}$.
In \cref{fig:tree_trajectories}, the edges of the main figure are an example of the posterior $p_\mathrm{S}(r_{1:t} \vert \phi_{1:t})$, inferred from the noisy images $\bm{\phi}_{1:t}$ that are represented by the edges of the inset.

For a given inferred path $r_{1:t}$, let $w(r_{1:t}, r^*_{1:t})$ be the fraction of time which $r_{1:t}$ is equal to the true path, $r^*_{1:t}$. Since on the tree, each path is equivalent to every other path, we can fix $r^*_{1:t}$ when calculating the overlap. The overlap is then
\begin{align} \label{eq:overlap}
    \overline{Y} & = \mathop{\mathbb{E}}_{r_{1:t}, \Phi} \left[ w(r_{1:t}, r^*_{1:t}) \right] \nonumber \\
    & = \mathop{\mathbb{E}}_{\Phi} \left[\frac{\sum_{r_{1:t}} w(r_{1:t}, r^*_{1:t}) e^{\epsilon_\mathrm{S} \Phi(r_{1:t})/\sigma_\mathrm{S}^2}}{Z_\mathrm{S}(\Phi)} \right].
\end{align}
When there is no signal $\epsilon_\mathrm{T}=0$, we expect all $k^t$ paths to be equally likely. Out of these paths, for a given overlap $wt$, there are $k^{t-1-wt}$ paths. Then the average overlap is a geometric sum,
\begin{align} \label{eq:overlap_unif}
    \overline{Y}_{\mathrm{unif}}(t) = \frac{1}{t} \sum_{wt=0}^{t} wt \, k^{-1-wt} \xrightarrow{t \gg 1} \frac{1}{t \ln k}.
\end{align}
Following the prescription in Eq. \eqref{eq:psi}, we separate the planted potential into iid Gaussian part and the signal part as $\phi_{x,t} = \psi_{x,t} + \epsilon_\mathrm{T} \delta_{x, x^*_t}$. Then, a path $r_{1:t}$ that has overlap $q$ with the true path $r^*_{1:t}$ has the energy $\Psi(r_{1:t}) + \epsilon_\mathrm{T} qt$, where $\Psi_{1:t}(r_{1:t}) = \psi_{r_1, 1} + \psi_{r_{1:2}, 2} + \cdots + \psi_{r_{1:t}, t}$ is the unplanted random field. Therefore we can write Eq. \eqref{eq:overlap} as an expectation on $\Psi$ as 
\begin{align} \label{eq:magnetisation}
    \overline{Y} = \mathop{\mathbb{E}}_\Psi \left[ \frac{\sum_w w z_w(\Psi, r^*) e^{\epsilon_\mathrm{S} \epsilon_\mathrm{T} wt / \sigma_\mathrm{S}^2}}{Z_{\mathrm{P}}(\epsilon_\mathrm{T}, \Psi)} \right],
\end{align}
where we introduced the partial partition function
\begin{align}
    z_w(\Psi, r^*) = \sum_{\substack{r_{1:t} \\ w(r_{1:t}, r^*_{1:t}) = w}} e^{\epsilon_\mathrm{S} \Psi(r_{1:t})/\sigma_\mathrm{S}^2},
\end{align}
and the total planted partition function now can be written as
\begin{align}
    Z_{\mathrm{P}}(\epsilon_\mathrm{T}, \Psi) = \sum_w z_w(\Psi, r^*) e^{\epsilon_\mathrm{S} \epsilon_\mathrm{T} wt / \sigma_\mathrm{S}^2}.
\end{align}
Then $Z_{\mathrm{P}}(0, \Psi)$ is the partition function for the directed polymer on a tree.

The directed polymer on a tree has been extensively studied, using the following three methods: the replica method, analogy to traveling waves \cite{derrida1988polymers, derrida1986solution}, and as a special case of the so-called generalised random energy model (GREM) \cite{derrida1986magnetic}. Below, we calculate the overlap between the true and inferred paths using the traveling wave and GREM approaches, and show that they lead to identical results.


\subsection{GREM approach}

The GREM was developed as an extension to the random energy model (REM), the simplest model for spin glass systems, to include correlations between states. 

Starting from the spin glass language, in the GREM, each Ising spin configuration $\sigma = (\sigma_1, \dots, \sigma_N)$ is assigned a path $r_{1:t}$ on a tree with total number of paths $k_1 \cdots k_t = 2^N$, with branching number $k_i$ generally varying with depth $i$. Then, each bond on the tree is assigned i.i.d Gaussian random energy with width $\sigma_i^2$ again varying with depth. Then, each spin configuration is given random but correlated energies corresponding to summing up energies of bonds for that path.

Assigning the ferromagnetic configuration $\sigma = (+1, \dots, +1)$ to the `true' path $r^*_{1:t}$, the overlap with the ferromagnetic configuration can loosely be interpreted as the magnetisation (per spin). Then, applying a magnetic field $b$ shifts the energies of the configurations with overlap $w$ by $w b N$.

In the case where with branching numbers and widths are constant with depth, the equations for the overlap with the true path from Bayesian inference and the magnetisation due to a magnetic field coincide and are given by Eq. \eqref{eq:magnetisation}, with inverse temperature $\beta := \epsilon_\mathrm{S}/\sigma_\mathrm{S}^2$ and magnetic field $b=\epsilon_\mathrm{T} t/N$. 



The case of the GREM in a magnetic field was studied by Derrida and Gardner in Ref. \cite{derrida1986magnetic}. There they argue that in the large $N$ limit, the quenched free energy can be approximated via `maximum a posteriori' (MAP) approximation. In our case, where we are interested in the large $t$ limit, this corresponds to
\begin{align}
    \lim_{t \rightarrow \infty} f = \max_{w} f_w,
\end{align}
where the quenched free energy (density) is $f := \frac{1}{t} \mathop{\mathbb{E}}_\Psi \left[ \ln Z_{\mathrm{P}}(\Psi) \right]$ and the partial free energy is 
\begin{align} \label{eq:partial_free_energy}
    f_w := \frac{1}{t} \mathop{\mathbb{E}}_\Psi \left[ \ln z_w (\Psi, r^*_{1:t}) \right] + \beta w \epsilon_\mathrm{T}.
\end{align}

The GREM undergoes phase transitions, the structure of which is given by the variation in $k_i$ and $\sigma^2_i$. In the constant $k_i=k$ and $\sigma^2_i=\sigma^2$ case, the GREM undergoes a single phase transition like the REM. 

Due to the structure of the tree, the partial partition function is another GREM \footnote{Note that the partial partition function will have a different branching number in the first timestep, leading to a correction term compared to when there is equal branching numbers. However, this correction is subextensive.}. The constant width and branching number case in the large $t$ limit can be solved \cite{derrida1988polymers} to be
\begin{align}
    \frac{1}{t} \mathop{\mathbb{E}}_\Psi \left[ \ln z_w(\Psi, r^*_{1:t}) \right] = (1-w) \beta c_\beta,
\end{align}
where the `speed of the front' $c_\beta$, whose naming will be made obvious later, is given by
\begin{align} \label{eq:speed_of_front}
    c_\beta = 
        \begin{dcases}
            c(\beta) & \text{if } \beta \leq \beta_c, \\ 
            c(\beta_c) & \text{if } \beta > \beta_c,
        \end{dcases}
\end{align}
with
\begin{align} \label{eq:speed_of_front_integral}
    c(\beta) = \frac{1}{\beta} \ln\left( k \int d \psi \rho_\mathrm{T}(\psi) e^{-\beta \psi} \right),
\end{align}
$\rho_\mathrm{T}(\psi) = \pi_\mathrm{T}(\psi) $ is the teacher's distribution of the iid noise, and the critical inverse temperature $\beta_c$ is given by $\frac{\partial}{\partial \beta} c(\beta) \rvert_{\beta = \beta_c} = 0$. The $w$-dependence of Eq. \eqref{eq:partial_free_energy} can be found to be
\begin{align} \label{eq:partial_free_energy_in_field}
    f_w = 
    \begin{dcases}
        \left(\beta \epsilon_\mathrm{T} - \left[\frac{\ln k}{\beta} + \frac{\beta \sigma_\mathrm{T}^2}{2} \right] \right) w + C_1 & \text{if } \beta \leq \beta_c, \\ 
        \left(\beta \epsilon_\mathrm{T}  - \sigma_\mathrm{T} \sqrt{2 \ln k} \right) w + C_2 & \text{if } \beta > \beta_c,
    \end{dcases}
\end{align}
where $C_1, C_2$ are constants independent of $w$ and $\beta_c = \sqrt{2 \ln k} / \sigma_\mathrm{T}$. As $f_w$ is linear in $w$, we see that $w$ that dominates the free energy is either $0$ or $1$, determined by the sign of the gradient term. The line where the gradient is zero then gives us the boundary between the two regimes, and can be found to be
\begin{align} \label{eq:phase_boundary}
    \frac{\epsilon_\mathrm{T}}{\sigma_\mathrm{T}} =
    \begin{dcases}
        \ln k \left( \frac{\epsilon_\mathrm{S} \sigma_\mathrm{T}}{\sigma_\mathrm{S}^2} \right)^{-1} + \frac{1}{2} \frac{\epsilon_\mathrm{S} \sigma_\mathrm{T}}{\sigma_\mathrm{S}^2} & \text{if  } \frac{\epsilon_\mathrm{S} \sigma_\mathrm{T}}{\sigma_\mathrm{S}^2} \leq \sqrt{2 \ln k}, \\ 
        \sqrt{2 \ln k} & \text{if  } \frac{\epsilon_\mathrm{S} \sigma_\mathrm{T}}{\sigma_\mathrm{S}^2} > \sqrt{2 \ln k}.
    \end{dcases}
\end{align}

\subsection{Traveling waves approach}
The overlap can also be calculated using the traveling waves approach. To do this we extend the arguments of \cite{derrida1988polymers} to the planted case. Using the structure of the tree, up to a multiplicative constant, the recursion relation for planted partition function can be written as
\begin{align}
    Z_{\mathrm{P}}(\epsilon_\mathrm{T}, t+1) = e^{\beta(\psi - \epsilon_\mathrm{T})} \left(Z^{(1)}_{\mathrm{P}}(\epsilon_\mathrm{T}, t) + \sum_{i=2}^{k} Z^{(i)}_{\mathrm{P}}(0, t)\right),
\end{align}
with the initial condition imposed to be $Z_{\mathrm{P}}(\epsilon_\mathrm{T}, 0) = 1$. We note that $\epsilon_\mathrm{T} = 0$ retrieves the unplanted partition function of the directed polymer. Defining the planted generating function as
\begin{align}
    G_{\epsilon_\mathrm{T}}(x, t) := \E_{\Psi} \left[ \exp(-Z_{\mathrm{P}}(\epsilon_\mathrm{T}, t) e^{-\beta x}) \right],
\end{align}
the recursion relation of the planted generating function can be written as
\begin{equation}
    \begin{aligned}
        G_{\epsilon_\mathrm{T}}(x,t+1) = \E_\psi \big[ & G_{\epsilon_\mathrm{T}}(x+\psi-\epsilon_\mathrm{T},t)  \\ \times \, & G_0(x+\psi -\epsilon_\mathrm{T},t)^{k-1} \big],
    \end{aligned}
\end{equation}
where $\psi \sim \mathcal{N}(0, \sigma_\mathrm{T}^2)$ represents noise at a single site/bond.

Let us consider continuous time evolution, where in a timestep $\delta t$, we can have branching with probability $\lambda \delta t$ (so $k=2$), and no branching with probability $(1-\lambda \delta t)$. We renormalise the noise to be $\psi \sim \mathcal N (0, 2 D \delta t)$ and the signal to be $\epsilon_\mathrm{T} \cdot 1 \rightarrow \epsilon_\mathrm{T} \delta t$. Then $G_{\epsilon_\mathrm{T}}$ evolves as
\begin{align} \label{eq:planted_wave}
    \partial_t G_{\epsilon_\mathrm{T}} = D \partial_x^2 G_{\epsilon_\mathrm{T}} - \epsilon_\mathrm{T} \partial_x G_{\epsilon_\mathrm{T}} - \lambda(1 - G_0) G_{\epsilon_\mathrm{T}}.
\end{align}
There are three components to Eq. \eqref{eq:planted_wave}. The first term proportional to $D$ is the diffusive term. The second term moves the planted generating function with characteristic velocity $\epsilon_\mathrm{T}$ towards the right. The third term is like the non-linear term in the Fisher-Kolmogorov-Petrovsky-Piskunov (FKPP) equation that pushes down $G_{\epsilon_\mathrm{T}}(x, t)$ at positions $x$ where $G_0(x, t) < 1$. 

It is known that the unplanted generating function $G_0(x, t)$ evolves according to the FKPP equation. In the long time limit, $G_0$ moves as a profile that switches from $0$ to $1$, and moves with speed $c_\beta$ given by Eq. \eqref{eq:speed_of_front}, $\lim_{t\rightarrow \infty} G_0(x,t) \approx w_\beta(x-c_\beta t)$, with $w_\beta(-\infty) = 0$ and $w_\beta(\infty)=1$.

\begin{figure}
    \centering 
    \includegraphics[width=0.87\linewidth]{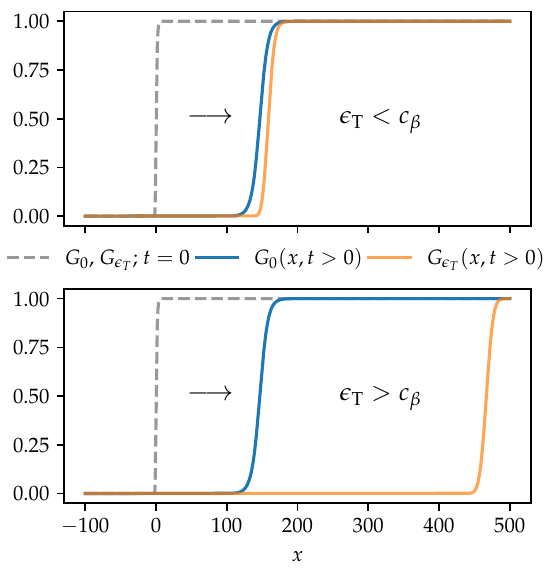}
    \caption{Numerical simulations of planted and unplanted generating functions. At $t=0$, the unplanted ($G_0$) and planted ($G_{\epsilon_\mathrm{T}}$) generating functions are equal (grey dashed line). The characteristic velocity of $G_{\epsilon_\mathrm{T}}$ is $\epsilon_\mathrm{T}$, while $G_0$ travels with $c_\beta$. (Top) $\epsilon_\mathrm{T} = c_\beta/2 < c_\beta$ for $t=60$. $G_{\epsilon_\mathrm{T}}$ (orange) sticks to $G_0$ (blue). (Bottom) As top but for $\epsilon_\mathrm{T} = 3 c_\beta /2 > c_\beta$. Here we see that $G_{\epsilon_\mathrm{T}}$ detaches from $G_0$, and moves with its characteristic velocity $\epsilon_\mathrm{T}$.} \label{fig:slow_and_fast} 
\end{figure}

\begin{figure}
    \centering 
    \includegraphics[width=\linewidth]{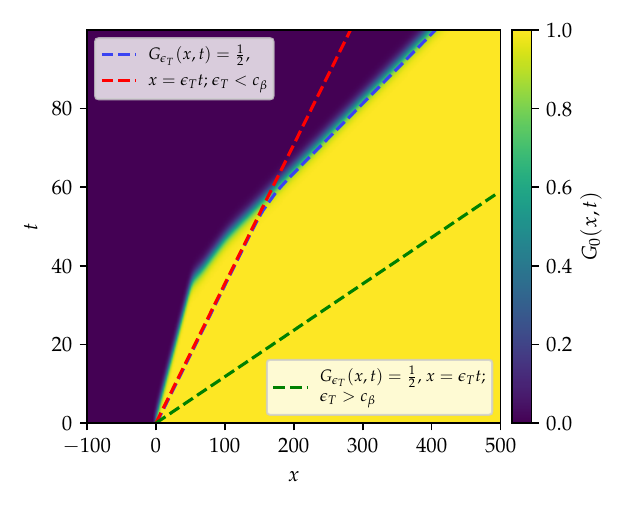}
    \caption{(Heatmap)
    Unplanted generating function ($G_0$) evolving according to the FKPP equation. (Blue dashed line) The front of the planted generating function ($G_{\epsilon_\mathrm{T}}$), initially moving with characteristic velocity $\epsilon_\mathrm{T}= c_\beta / 2$ (red dashed line) smaller than the unplanted velocity $c_\beta$. The front of $G_{\epsilon_\mathrm{T}}$ eventually sticks to that of $G_{0}$. (Green dashed line) For characteristic velocities larger than $c_\beta$ ($\epsilon_{\mathrm{T}} = 3 c_\beta / 2 > c_\beta$), $G_{\epsilon_\mathrm{T}}$ is free to travel at $\epsilon_\mathrm{T}$.
    } \label{fig:generating_functions}
\end{figure}

$G_0$ and $G_{\epsilon_\mathrm{T}}$ start with the same initial conditions. If $\epsilon_\mathrm{T} < c_\beta$, then the third term determines the position of the front and $G_{\epsilon_\mathrm{T}}$ moves with the same velocity as $G_0$. If $\epsilon_\mathrm{T} > c_\beta$, then the support of $G_{\epsilon_\mathrm{T}}$ moves to a region where $G_0 = 1$ and therefore the third term vanishes, leading to a linear PDE for an error function that moves towards the right with speed  $\epsilon_\mathrm{T}$ and spreads as $\sqrt{t}$. Therefore $G_{\epsilon_\mathrm{T}}$ moves with velocity $v_{\beta, \epsilon_\mathrm{T}} = \max(\epsilon_\mathrm{T}, c_\beta)$. This is supported and illustrated with numerical simulations in \cref{fig:slow_and_fast,fig:generating_functions}.

Now $G_{\epsilon_\mathrm{T}}(x, t)$ switches from $0$ to $1$ approximately at point $\beta \hat x \sim \ln Z_{\mathrm{P}}(\epsilon_\mathrm{T}, t)$. From the previous arguments, this point at long times is $\hat x(t) = v_{\beta, \epsilon_\mathrm{T}} t$. Assuming that fluctuations in $\ln Z_{\mathrm{P}}(\epsilon_\mathrm{T}, t)$ are sublinear, this means that $\E_\Psi \left[ \ln Z_{\mathrm{P}}(\epsilon_\mathrm{T}, t)\right] = \beta v_{\beta, \epsilon_\mathrm{T}} t$.

    
From the definition of average overlap $Y$, we have that $Y = \frac{1}{\beta t} \frac{\partial}{\partial \epsilon_\mathrm{T}} \E_\Psi \left[ \ln Z_{\mathrm{P}} (\epsilon_\mathrm{T}, t)\right] = \frac{\partial}{\partial \epsilon_\mathrm{T}} v_{\beta, \epsilon_\mathrm{T}}$. Then, when $\epsilon_\mathrm{T} \leq c_\beta$, $v_{\beta, c_{\epsilon_\mathrm{T}}}$ does not change with $\epsilon_\mathrm{T}$ and therefore $Y=0$. On the other hand, when $\epsilon_\mathrm{T} > c_\beta$, then we have $\partial_{\epsilon_\mathrm{T}} v_{\beta, c_{\epsilon_\mathrm{T}}} = 1$ and therefore have the overlap of $Y=1$. As $c_\beta$ has a phase transition, we must consider the cases when $\beta \leq \beta_c$ and $\beta > \beta_c$ after which we retrieve the identical phase boundary Eq. \eqref{eq:phase_boundary}.

\subsection{General noise distribution}

We can generalise to the case where the teacher uses an arbitrary noise distribution, whilst the student conducts inference assuming Gaussian noise. This can be done by choosing a different distribution for $\rho_\mathrm{T}(\psi)$ in \cref{eq:speed_of_front_integral}. The minimum speed is again given by $\partial_\beta c(\beta)\rvert_{\beta=\beta_c}=0$. For example, for the top-hat distribution $\rho_{\mathrm{T}}(\psi) = 1/2a$ for $-a<\psi<a$, we have $c(\beta) = \ln \left( k \sinh(\beta a)/\beta a\right)/\beta$, with minimum given at $\beta_c = x(k)/a$ for $x\coth(x)-\ln(k\sinh(x)/x)=1$. This is the same qualitative picture as \cref{fig:phase_diagram}, where $c_\beta$ decreases until it hits minimum at some $\beta_c$ then freezes. Numerically testing on various unimodal distributions, we consistently found the same qualitative picture.

\subsection{Phase diagram and numerics}
\begin{figure}[h]
\center 
\begin{tikzpicture}[scale=1.5, samples=100]
    \draw[->] (0, 0) -- (3, 0) node[right] {$\epsilon_\mathrm{S} \sigma_\mathrm{T}/\sigma_\mathrm{S}^2$};
    \draw[->] (0, 0) -- (0, 3) node[above] {$\epsilon_\mathrm{T}/\sigma_\mathrm{T}$};
    \draw[-] (1, 0) node[below] {$\sqrt{2 \ln k}$};
    \draw[-] (0, 1) node[left] {$\sqrt{2 \ln k}$};
    \draw[-] (0, 0) node[below left] {$0$};
    \draw[-] (2.0,2.0) node[] {$Y=1$};
    \draw[-] (2.0,0.5) node[] {$Y=0$};
    \draw[smooth, domain = 0.175:1, color=blue] plot (\x,{1/\x/2 + \x/2});
    \draw[smooth, domain = 1:3, color=blue] plot (\x,{1});
    \draw[smooth, domain = 0:1, color=blue, dashed] plot (1,\x);
    \draw[smooth, domain = 0:1, color=blue, dashed] plot (\x,1);
\end{tikzpicture}
\caption{Phase diagram for overlap $Y$ on the plane of teacher-student signal strengths $\epsilon_{\mathrm{S}}-\epsilon_{\mathrm{T}}$ for the case of a tree with branching number $k$. $Y=1$ corresponds to perfect inference whilst $Y=0$ corresponds to failure of inference. The boundary corresponds to a 1st order phase transition.}
\label{fig:phase_diagram}
\end{figure}
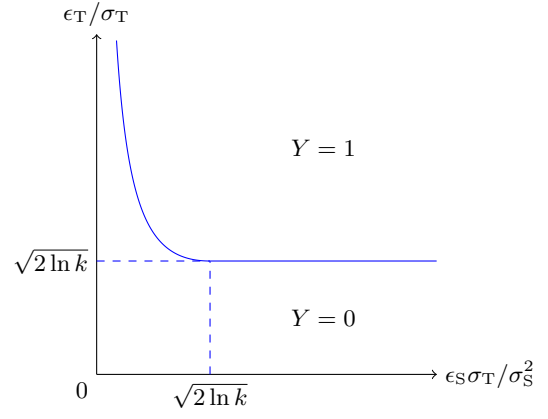

The phase diagram obtained via either approach with boundary given by Eq. \eqref{eq:phase_boundary} is sketched in \cref{fig:phase_diagram}.

In the Bayes optimal case, corresponding to the $45^{\circ}$ line on \cref{fig:phase_diagram}, the planted directed polymer `inherits' the unplanted model's transition point, $\epsilon/\sigma = \sqrt{2 \ln k}$. Additionally, for a given $\epsilon_\mathrm{T}/\sigma_\mathrm{T}$, one cannot get a greater overlap than the Bayes optimal case. \changes{This is expected as it is a standard result that one cannot do better than the Bayes optimal case for various tasks such as classification and estimation \cite{bishop2006pattern, pishro2014introduction}. However, for the specific case of the tree, it appears that as long as $\epsilon_\mathrm{S} > \sqrt{2 \ln k}$, performance in inference is only limited by $\epsilon_\mathrm{T}$. This suggests that in the case where the student does not know the true value of the signal strength, they can pick any value $\epsilon_\mathrm{S} > \sqrt{2 \ln k}$ and will be able to infer as well as the Bayes optimal case. 
A separate partially unanswered question is how sure the student is about their answer, that is the variance or uncertainty on their inferred path. This could be quantified by overlap between two paths $r^{(1)}_{1:t}$, $r^{(2)}_{1:t}$ sampled from the student's posterior,
\begin{equation}
    \overline{Y_\mathrm{S}} := \frac{1}{t} \E_{\bm{y}_{1:t}} \left[\E_{r^{(1)}_{1:t}, r^{(2)}_{1:t} \sim p_\mathrm{S}(\cdot \vert \bm{y}_{1:t})} \left[ \sum_{\tau=1}^t \delta_{r^{(1)}_\tau, r^{(2)}_\tau} \right] \right].
\end{equation}
The case of zero teacher SNR ($\epsilon^*=0$) corresponding to the usual directed polymer case was studied in \cite{derrida1988polymers}, which revealed a transition from $\overline{Y_\mathrm{S}}=0$ to $1$ at $\epsilon_\mathrm{S} \sigma_\mathrm{T} / \sigma_\mathrm{S}^2 = \sqrt{2 \ln k}$, coincides with our transition point at Bayes optimality. The symmetry between the student and teacher means that $\overline{Y_\mathrm{S}}=\overline{Y}$ in the Bayes optimal case, which immediately gives us $\overline{Y_\mathrm{S}}$ along the line $\epsilon_\mathrm{T}/\sigma_\mathrm{T} = \epsilon_\mathrm{S} \sigma_\mathrm{T} / \sigma_\mathrm{S}^2$ in \cref{fig:phase_diagram}. 
This agrees with the transition in the averaged entropy of the posterior, in the Bayes optimal case found in the recent preprint Ref. \cite{gerbino2024measurement} for a generalised measurement model. 
It would be interesting to extend either of these quantities to the rest of the phase diagram.
}

\changes{The reason for the first order `all-to-nothing' transition in the tree case is due to the fact that in the thermodynamic limit all overlap ($w$) dependent contributions to the free energy are linear. This is because any path with overlap $w$ is a combination of \emph{the} path following the planted trajectory (with purely energetic and no entropic contribution to the free energy) followed by a path which has left it, never to return, and is therefore accounted for by the solution of the unplanted problem. The transition between $w=0$ and $w=1$ is therefore determined purely by the sign of the linear term. This is not the case in the finite dimensional problem on a lattice, for example.}

\begin{figure}
    \centering
    \includegraphics[width=\linewidth]{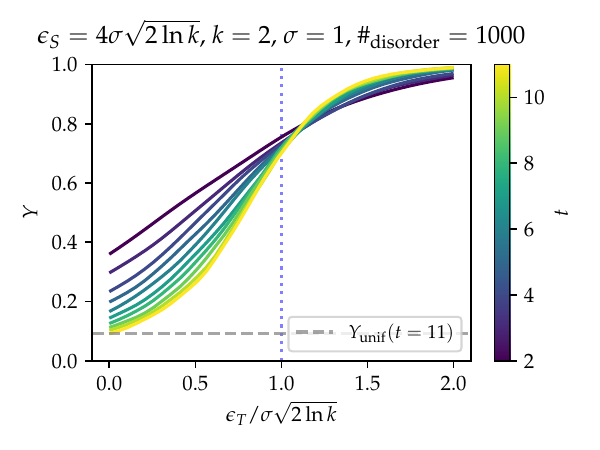}
    \caption{Numerical study of the average overlap $Y$ with teacher signal strength $\epsilon_{\mathrm{T}}$ at various times $t$ at fixed branching number $k=2$ and student signal strength $\epsilon_\mathrm{S} = 4 \sigma \sqrt{2 \ln k}$. $\#_{\text{disorder}}=1000$ is the number of disorder realisations.}
    \label{fig:tree_overlap_numerics}
\end{figure}

To confirm our theory, in \cref{fig:tree_overlap_numerics}, we numerically study the overlap as we vary time $t$ up to $t=11$ for fixed $\sigma_\mathrm{S} = \sigma_\mathrm{T} = \sigma = 1$, $k=2$, and $\epsilon_\mathrm{S} = 4 \sigma \sqrt{2 \ln k}$ for $\#_{\mathrm{disorder}}=1000$ disorder realisations. We confirm the `finite size scaling' of Eq. \eqref{eq:overlap_unif} for $\epsilon_\mathrm{T}=0$, and see that the numerics support theoretical transition point $\epsilon^{*}_{T} = \sigma_\mathrm{T} \sqrt{2 \ln k}$.

\section{Conclusion} \label{sec:conclusions}

In this work we introduced the planted directed polymer problem. \changes{Our main contributions are as follows: 1) we derived an explicit mapping of the inference problem to the directed polymer. 2) We made an explicit connection to the statistical physics of inference in the planted ensemble. 3) We solved the planted polymer on a tree exactly, adapting existing machinery developed for the unplanted case. 4) We provide heuristic arguments combining numerical and analytical reasoning for the absence of a phase transition in the one dimensional problem, as well as discussing qualitiative features of the free energy profile and the higher dimensional versions of the problem.}

Earlier work of Yuille and Coughlan \cite{yuille2000fundamental} studied tracking of roads in aerial images as a problem of Bayesian inference, including the notion of the planted ensemble. They raised the question of whether there is a fundamental limit to the success of inference in this situation and connected this limit to a phase transition in the thermodynamic limit. They made the further simplification of restricting the paths to a tree, as we did in Section \ref{sec:tree}, but without connecting to the earlier literature on the directed polymer. 
They obtain a bound on the `error' $t (1-Y)$ that depends on a regularisation parameter used to treat the $t\rightarrow \infty$ limit but not on $t$. Therefore the regime of vanishing overlap $Y=0$ is not handled correctly.
In Ref.~\cite{yuille1999high} the same authors explored the effect of non-optimality due to inference based on the wrong prior.


More recent work of Offer \cite{offer2018phase} addressed the related problem of Bayesian tracking in \emph{clutter} using a mean field approximation as well as numerical evaluation of the posterior, finding a phase transition in three or more spatial dimensions. Again, the connection to the directed polymer was not made. 

\changes{A separate but related question is whether or not a random walker exists in the images \emph{at all}. Refs. \cite{burnashev1982statistical,agaskar2015optimal} addressed this question by testing the hypothesis of there being a random walker versus the null hypothesis of no walker (i.e. images of white noise). They relate the decay rate of the false negative probability to the average log-likelihood ratio between the hypothesis and null hypothesis. The log-likelihood ratio coincides with the free energy of the directed polymer. It would be interesting to apply the known results of the directed polymer literature to this setting.}

Several other avenues for future research exist. One could attempt to adapt existing analytical tools for the $d=1$ directed polymer (such as the replica Bethe ansatz, see e.g. Ref.~\cite{le2012kpz}) to the planted case in order to calculate some of the observables explored in this paper. Moving to higher dimensions, it is known that the directed polymer has a transition at finite disorder strength for $d \geq 3$. We may expect the planted directed polymer to `inherit' the phase transition, as suggested by the investigation of Ref.~\cite{offer2018phase}.



\subsection*{Acknowledgements}
SWPK and AL acknowledge support from EPSRC DTP International Studentship Grant Ref. No. EP/W524475/1 and EPSRC Critical Mass Grant EP/V062654/1 respectively. We are grateful to Bernard Derrida for useful discussions.





\bibliographystyle{apsrev4-2} 
\bibliography{bibliography} 





\end{document}